\documentclass[10pt,amsmath,amssymb,nofootinbib,twoside,twocolumn,superscriptaddress,floats,floatfix,aps,prd,preprintnumbers]{revtex4-2}

\usepackage[british]{babel}
\usepackage{xcolor}
\usepackage{booktabs}
\usepackage{latexsym}
\usepackage{amssymb}
\usepackage{amsmath}
\usepackage{graphicx}
\usepackage{ dsfont }
\usepackage{mathrsfs} 
\usepackage{tabularx}
\usepackage{cancel}
\usepackage{hyperref}
\hypersetup{
    colorlinks=true,
    linkcolor=black,
    citecolor=blue,
    filecolor=black,
    urlcolor=blue,
    pdfauthor={M. Cadoni and A. P. Sanna},
    pdftitle={Nonsingular black holes from conformal symmetries}
}
\usepackage[capitalise]{cleveref}
\usepackage[utf8]{inputenc}
\usepackage[normalem]{ulem}
\def\beq#1\eeq{\begin{align}#1\end{align}}

\newcommand{\dd}{\text{d}}

\newcommand{\ellP}{\ell_\text{P}}

\begin{document}

\newcommand{\UniCa}{\affiliation{Dipartimento di Fisica, Universit\`a di Cagliari, Cittadella Universitaria, 09042 Monserrato, Italy}}
\newcommand{\INFNCa}{\affiliation{INFN, Sezione di Cagliari, Cittadella Universitaria, 09042 Monserrato, Italy}}

\author{M.~Cadoni}
\email{mariano.cadoni@ca.infn.it}
\UniCa\INFNCa

\author{A.~P.~Sanna}
\thanks{Corresponding author}
\email{asanna@dsf.unica.it}
\UniCa\INFNCa

\title{Nonsingular black holes from conformal symmetries}

\begin{abstract}
We derive the form of the metric for static, nonsingular black holes with a de Sitter core, representing a deformation of the Schwarzschild solution, by assuming that the gravitational sources  describe a flow between two conformal points, at small and great distances. The resulting black-hole metric turns out to be a particular case of the Fan $\&$ Wang metric, whose parameters have been recently constrained by using the data  of the S$2$ star orbits around the galactic centre SgrA$^\ast$.   
\end{abstract}

\maketitle

\section{Introduction}
\label{sec:introduction}

In recent times there has been renewed interest for asymptotically flat, nonsingular black-hole solutions, which deform the Schwarzschild solution at subleading order \cite{bardeen1968proceedings,Dymnikova:1992ux,Bonanno:2000ep,Dymnikova:2004qg,Hayward:2005gi,Nicolini:2005vd,Ansoldi:2008jw,Fan:2016hvf,Frolov:2016pav,Simpson:2019mud,Simpson:2018tsi,Carballo-Rubio:2019fnb,Cadoni:2022chn,Sebastiani:2022wbz,Cadoni:2022vsn,Akil:2022coa}. Among them, the most interesting class of solutions is represented by nonsingular black holes with a de Sitter (dS) core \cite{Cadoni:2022chn}. 
These black-hole solutions are of interest for several reasons. Firstly, they allow to circumvent Penrose's theorem \cite{Penrose:1964wq} by  removing the classical singularity at $r = 0$. 
Secondly, they are solutions of Einstein's field equations sourced by an anisotropic fluid, effectively encoding the deviations responsible for the smearing of the singularity. These deviations are described by an external length scale $\ell$, which represents an additional ``hair" of the black hole. An intriguing possibility is that it could be also of superplanckian  origin \cite{Cadoni:2022chn}. Thirdly, they can play the role of black-hole ``mimickers", i.e., they  are indistinguishable from the  Schwarzschild solution at great distances, but could nonetheless lead to observable deviations from the latter, for instance in the orbits of massive particles and photons and in the gravitational-wave spectrum (see Refs.~\cite{Cadoni:2022chn,Cadoni:2022vsn} and references therein).
Last but not least, they could be very useful in solving the information puzzle  arising during black-hole evaporation \cite{Cadoni:2023tse}.

On the other hand, such models suffer from a strong limitation, which is purely theoretical. We can obtain them using general relativity (GR) with anisotropic fluids as sources, but  the underlying microscopic physics is mostly unknown. This difficulty becomes  particularly severe in those cases in which the deformations from the usual Schwarzschild solution have   superplanckian origin \cite{Cadoni:2022chn}. 
The consequence is that we have a huge degeneracy, giving rise to a  broad  class of metric solutions, which all describe nonsingular black holes with a dS core.
The coarse-grained description in term of the anisotropic fluid is not stringent. The  equation of state (EOS) relating the radial pressure with the energy density, $p_\parallel = p_\parallel(\rho)$, and the density profile $\rho(r)$, interpolating between small and large $r$, are very weakly constrained.   

Probably these difficulties are hinting at the fact that the microscopic explanation of this kind of solutions cannot be found by merely looking at GR, which only allows for an effective description of the sources in terms of anisotropic fluids. What is needed, then, is a general guiding principle to select the physically relevant solutions. 

In this paper, we will use conformal symmetries, including Lorentz boosts, as the guiding principle to remove the above mentioned degeneracy of solutions.   
There is striking evidence that conformal symmetry could be a crucial feature of any quantum theory of gravity. It is the pillar of the AdS/CFT correspondence  \cite{Maldacena:1997re,Witten:1998qj} and is also crucial for most microscopic derivations of the Bekenstein-Hawking black-hole entropy \cite{Strominger:1996sh,Strominger:1997eq,Cadoni:1998sg,Cadoni:1999ja,Carlip:2002be}.
 
We will select the EOS for the anisotropic fluid using covariance under rotations and radial Lorentz boosts.
In order to fix the density profile $\rho(r)$, we will use conformal symmetries, motivated by the role played by the latter in black-hole physics. In particular, we will assume that 
$\rho(r)$ describes the flow of matter fields between two conformal points where the local matter contribution is negligible, near $r=0$ and at $r = \infty$. These are described, respectively, by the dS spacetime \cite{Keane:1999kp,Cadoni:2006ww} and a CFT in Minwkoski spacetime. The existence of the latter, in particular, represents a strong assumption, which is, however, motivated by the AdS/CFT correspondence.
We will show that these requirements select a specific spacetime metric, i.e., a particular case of the Fan $\&$ Wang metric \cite{Fan:2016hvf}, which represents the one having the strongest subleading deviations from Schwarzschild at infinity and which was recently constrained by S$2$ observational data \cite{Cadoni:2022vsn}. 

The present paper is organized as follows. 

In \cref{sec:nonsingular}, we will briefly review some basic features of nonsingular black holes with a dS core and we will fix the EOS using Lorentz symmetries. In \cref{sec:scalingssource}, we discuss the conformal symmetries we use to constrain the density profile and derive the form of the metric. In \cref{sec:nle}, we give a simple example for the source in terms of nonlinear electrodynamics. Finally, in \cref{sec:concl} we summarize our results.

\section{Nonsingular black holes with a de Sitter core}
\label{sec:nonsingular}

Due to Birkhoff's theorem, any nonstandard GR black-hole solution has to be obtained from Einstein's equations sourced by a nonzero stress-energy tensor. The most general one is that of an anisotropic fluid, which has been widely adopted to effectively parametrize several different effects and deviations from GR phenomenology, both at small and cosmological scales (for an incomplete list, see, e.g., Refs.~\cite{Bayin:1985cd,Herrera:1997plx,DeBenedictis:2005vp,Aluri:2012re,Raposo:2018rjn,Cadoni:2017evg,Cadoni:2020izk,Cadoni:2022chn,Simpson:2019mud,Simpson:2018tsi,Akil:2022coa} and references therein). This fluid is described by the stress-energy tensor\footnote{Throughout the entire paper, we will use natural units in which $\hbar = c = 1$ and we will use $G = \ellP^2$ interchangeably. }
\begin{equation}
    T_{\mu\nu} = \left(\rho + p_\perp \right)u_\mu u_\nu + p_\perp g_{\mu\nu} + \left(p_\parallel-p_\perp \right)w_\mu w_\nu\, ,
    \label{AnistropicSEI}
\end{equation}
where $\rho$, $p_\parallel$ and $p_\perp$ are the energy density, the radial and perpendicular components of the pressure, respectively, while $u_\mu$ and $w_\mu$ are a time-like and space-like 4-vectors, respectively, satisfying the relations $u^\mu u_\mu = -w^\mu w_\mu = -1$. A particular choice of the EOS $p_\parallel = p_\parallel(\rho)$ determines and characterizes the solutions, whereas $p_\perp$ is determined by the covariant conservation of the stress-energy tensor (see, e.g., Refs.~\cite{bardeen1968proceedings,Hayward:2005gi,Dymnikova:1992ux,Chan:2011ayt,Beltracchi:2018ait,Simpson:2019mud,Fan:2016hvf,Cadoni:2022chn,Akil:2022coa,Nicolini:2005vd,Kumar:2021oxa} and references therein). 

Requiring symmetry properties of the fluid constrains the free functions $\rho$ and $p_\parallel$ in \cref{AnistropicSEI}. In the following, we focus on spherically-symmetric models. Given this, we consider fluids whose dynamic equations are covariant under rotations in the $[\theta, \phi]$ plane and under Lorentz boosts in the $[t, r]$ directions. Even if every stress-energy tensor is covariant under boosts and rotation in general frames, these choices select invariance for the particular class of radially-moving observers. The physical consequence of this choice is that a stress-energy tensor satisfying these properties is identified as describing a well-defined spherically symmetric and Lorentz invariant vacuum \cite{gliner1966algebraic,Dymnikova:1992ux,Dymnikova:2004qg}. Its structure reads
\begin{subequations}
    \begin{align}
        &T^\theta_\theta = T^\phi_\phi\, ; \label{sphericalpart}\\
        &T^t_t = T^r_r \, . \label{radialpart}
    \end{align}
\end{subequations}
\Cref{radialpart}, in particular, fixes the EOS to be
\begin{equation}
    p_\parallel = -\rho\, .
    \label{EoS}
\end{equation}
Notice that, apart from being dictated by symmetry arguments, this EOS is quite natural, as well as simple, in an emergent gravity framework (see, e.g., Refs.~\cite{Cadoni:2022chn,Jusufi:2022cfw}). Additionally, it appears in several  physical contexts, such as the simplest form of dark energy (the cosmological constant, in the isotropic case), exotic compact objects \cite{MartinMoruno:2011rm,Beltracchi:2018ait} or solutions of GR coupled with nonlinear electrodynamics \cite{Bronnikov:2000vy,Dymnikova:2004zc}. 

$p_\perp$, instead, is entirely determined by the covariant conservation of the stress-energy tensor. 
\begin{equation}
   p_\perp = -\rho -\frac{r}{2}\rho'\, .
    \label{pperp}
\end{equation}
With \cref{EoS}, the general solution of Einstein's equations, sourced by \cref{AnistropicSEI} and written in Schwarzschild coordinates $(t, r, \theta, \varphi)$, reads (see, e.g., Ref.~\cite{Cadoni:2022chn})
\begin{subequations}
\begin{align}
    &\dd s^2 =-f(r) \dd t^2 + \frac{\dd r^2}{f(r)} + r^2 \dd \Omega^2\, ; \label{generalmetric}\\
    &f(r) = 1- \frac{2 G m(r)}{r}\, , \quad m(r) = 4\pi \int_0^r \dd \tilde r \, \tilde r^2 \, \rho(\tilde r)\, ,
    \label{MSmassandf}
\end{align}
\end{subequations}
where $m(r)$ the Misner-Sharp mass of the system. 

Note that we could have adopted a different parametrization of the metric \eqref{generalmetric}, resulting therefore in a different definition of the radial coordinate. For example, the line element written in isotropic coordinates
\begin{equation}
    \dd s^2 = -A(r') \dd t^2 + B(r')\left(\dd r'^2 + r'^2 \dd \Omega^2 \right)
\end{equation}
is related to \cref{generalmetric} by the coordinate transformation $\dd r'/r' = \dd r/\left(r \sqrt{f} \right)$, relating the two radial coordinates. All the relevant physical results  of the present paper are essentially independent of the particular parametrization  of the radial coordinate. The freedom in the choice of the latter would simply amount to a different realization of the same symmetries analyzed below, which, depending on the chosen coordinates, could have an intricate form. In what follows, therefore, we will limit ourselves to considering the parametrization \eqref{generalmetric}, which allows for a simple realization of such symmetries and is the one most widely used in the literature.
\\

Considering, thus, the system \eqref{generalmetric}-\eqref{MSmassandf}, \cref{EoS} fixes $p_\parallel$ and $p_\perp$, but leaves the density profile and, hence, the form of $m(r)$,  completely unconstrained. 

On the other hand, the behavior of $\rho$ at $r=0$ and $r \to \infty$ can be determined by stringent physical considerations.

In light of the particular form of the EOS \eqref{EoS} and requiring the absence of spacetime singularities, we expect that, whenever matter contribution is negligible (at $r \sim 0$ and $r \to \infty$), the source of gravity is given by an approximately isotropic fluid, which gives $\rho \sim \text{constant}$ using \cref{EoS,pperp}. Assuming the validity of the weak energy condition we have  $\rho \geq 0$. From \cref{EoS}, it follows now that, in the  core, at $r \sim 0$, the spacetime behaves as a dS spacetime \footnote{See, however, Ref.~\cite{Simpson:2019mud} for a model with an asymptotic Minkowski core.}. The density reads
\begin{equation}
    \rho \sim \frac{1}{4\pi \ellP^2 \, L^2}\, ,
    \label{rhodS}
\end{equation}
where $L$ represents the dS length in the core. This behavior at the center breaks the strong energy condition, allowing to circumvent Penrose's theorem and to replace the classical singularity region with a completely regular spacetime \cite{bardeen1968proceedings,Hayward:2005gi,Dymnikova:1992ux,Fan:2016hvf,Cadoni:2022chn,Nicolini:2005vd}. \Cref{rhodS} constrains the Misner-Sharp mass \eqref{MSmassandf} to behave extensively, as $m(r) \sim r^3/\ellP^2 L^2$, near  $r= 0$. 
  
If one considers the cosmological regime, dominated by a cosmological constant at large  $r$, one can still have $\rho \sim \text{constant} \neq 0$, and hence a dS behavior.
Since we are considering isolated bodies, we discard a dS asymptotics, assuming that the density profile decays sufficiently rapidly to zero at $r \to \infty$, so that we have asymptotically flat solutions. We also note that, at infinity, $p_\parallel = p_\perp \to 0$ according to \cref{pperp}. 

Moreover, the asymptotic value $M$ of $m(r)$ appears as an integration constant in \cref{MSmassandf}, so that imposing a Schwarzschild  behavior for the solution at $r \to \infty$ implies 
\begin{equation}
\label{asrho}
\rho\sim r^c\,, \quad \text{with} \quad  c<-3\, .
\end{equation}
Thus, our physically motivated ``boundary conditions" at $r=0$ and $r\to\infty$ imply that the function $\rho$ interpolates between the constant value \eqref{rhodS} near $r=0$ to \cref{asrho} at $r \to \infty$.

A major drawback of this construction, however, is that, as a consequence of the freedom in choosing both the exponent $c$ in \cref{asrho} and the interpolating density profile $\rho(r)$, the model  is not unique, but it exists an infinite class of models which realize the same flow \cite{Cadoni:2022chn}. Particularly relevant examples, discussed in the literature, are the Fan $\&$ Wang \cite{Fan:2016hvf}, Bardeen \cite{bardeen1968proceedings}, Hayward \cite{Hayward:2005gi} nonsingular black holes, and black holes with Gaussian cores \cite{Bonanno:2000ep,Nicolini:2005vd} (see also Ref.~\cite{Cadoni:2022chn}  and references therein).

In the following, we will see that requiring some conformal symmetries and scaling properties for the density profile and for the field generating this energy density will select a particular metric belonging to this general class.  

\section{Conformal symmetries  and scalar field description }
\label{sec:scalingssource}

Although GR is not a conformal field theory (see, however, Ref.~\cite{Cadoni:2006ww}), it is known that these  symmetries could play an important role for particular spacetime backgrounds, like, e.g., the anti de Sitter (or also the dS) spacetime, for which they take the form of holographic  correspondences \cite{Maldacena:1997re,Witten:1998qj,Strominger:2001pn}.
Moreover, there is some evidence that conformal symmetry could regularize   the short-distance behavior of gravity, by generating an  UV fixed point, which is at the base of the asymptotic safety scenario \cite{Bonanno:2020bil}.

Conformal symmetry plays also an important role for black holes, in particular in the description of their near-horizon physics. It has been widely used to give a microscopic derivation of the Bekenstein-Hawking entropy   \cite{Strominger:1996sh,Strominger:1997eq,Cadoni:1998sg,Carlip:2002be}. Moreover, extremal black-hole background geometries (e.g., BPS states) typically describe the flow between different conformal points, or a conformal point and a flat spacetime \cite{Ferrara:1996dd,Gubser:2008wz}.

Finally, conformal symmetries are very important also for nonsingular black holes with a dS core. In fact, the $4$D dS spacetime is endowed with a scale invariance  \cite{Cadoni:2006ww} and, in particular, invariance under transformations induced by the conformal group SO$(2,4)$ \cite{Keane:1999kp} \footnote{This becomes evident when embedding dS spacetime in $\mathds{R}^{1,4}$ and writing it in the flat slicing.}. For the nonsingular black holes under consideration here, this scale symmetry holds only in the dS core and it is broken at greater distances, when localized matter begins to dominate  \cite{Cadoni:2022chn}. 
On the other hand, the presence of the dS core implies that, for some values of the hair $\ell$,   the black hole has necessarily two horizons, which, for a critical value of $\ell$, merge into a single one. This produces an extremal configuration, whose near-horizon geometry has an AdS$_2$ factor, with an associated dual, near-horizon conformal symmetry \cite{Cadoni:2022chn}.

These considerations strongly suggest that the density  $\rho(r)$ sourcing our black hole could generate a flow between a conformal point near $r=0$, described by the dS spacetime, and some conformal field theory in the $r =\infty$ region.
The scale invariance is broken  during the flow by the nucleation of a local mass $M$, with a related generation of an intermediate scale $\ell$ \cite{Cadoni:2022chn}. The latter represents an additional ``hair" of these models, and allows to realize the interpolation between the small $r \sim 0$ and the large scales $r \to \infty$.

Lacking a fundamental microscopic description of our nonsingular black holes, we are unable to exactly identify the field content of the conformal matter sourcing the black hole in the $r\to \infty$ region. However, scale symmetry strongly constraints the form of $\rho$ in this regime. It must transform with definite weight $\Delta$ under dilatations $r \to \omega r$: $\rho(\omega  r) = \omega^\Delta \rho(r)$.   For conformal field theories, the scaling dimension $\Delta$ must be equal to the engineering dimensions, $\Delta =-4$, in such a way that the theory does not contain dimensional constants. This fixes the exponent $c$ in the asymptotic behavior \eqref{asrho}, so that we have  
\begin{equation}
    \rho(r) = \frac{\alpha}{4\pi} \frac{1}{r^4}\, ,
    \label{rhodS1}
\end{equation}
where $\alpha$ is a dimensionless constant. This scaling is typical of the energy density of conformal matter  fields in four dimensional Minkowski spacetime, like, for instance, a set of free masseless scalar fields.  
 We are therefore assuming that, in the asymptotic $r=\infty$ region, if we neglect the contribution of the localized matter $M$, our system is well described by a solution of GR given by a CFT in Minkowski spacetime, whose energy density corresponds to \cref{rhodS1}.
Na\"ively, this energy density characterizes a system of $N$ quanta inside a sphere of radius $r$. Each mode has a typical Compton energy $E \sim r^{-1}$, so that the total energy density is $\rho \sim N/(r \cdot r^3)= N/r^4$.

The density profile \eqref{rhodS1} diverges at $r=0$. This is due to modes with arbitrarily short wavelength contributing to the spectrum. According to our assumption on the presence of a dS behavior at $r = 0$, this singularity is, however, not physical, because the density $\rho$ must interpolate between the constant value \eqref{rhodS} at $r=0$ and \cref{rhodS1} at great distances.

The simplest way to regularize this divergent behavior is through a translation of the radial coordinate $r \to r+ \ell$, which moves the singularity to nonphysical negative values of the radial coordinate $r$
\begin{equation}
    \rho(r) = \frac{\alpha}{4\pi} \frac{1}{(r+\ell)^4}\, .
    \label{rhodS2}
\end{equation}
This introduces a length scale related to the local mass $M$  and  to the dS length $L$, which is the physical source of the breaking of the scale symmetry. Evaluating \cref{rhodS2} in $r=0$, comparing it with \cref{rhodS} and considering the Schwarzschild limit $m(r) \to M$ as $r \to \infty$, we can easily identify the dimensionless constant $\alpha$ and write the   hair $\ell$ in terms of the Schwarzschild radius $R_\text{S} = 2 \ellP^2 M$  and of $L$ 
\begin{equation}
  \alpha = \frac{\ell^4}{\ellP^2 \, L^2}\,,\quad  \ell \sim R_\text{S}^{1/3} \, L^{2/3}\, .
    \label{ellRS}
\end{equation}
The second equation, in particular, gives a universal scaling for every geometry interpolating between the dS at $ r = 0$ and the Schwarzschild spacetimes at great distances (see Ref.~\cite{Cadoni:2022chn}). Specifically, \cref{ellRS} represents a universal relation between $\ell$ and the black-hole mass $M$.

One can now easily find the mass function, using \cref{rhodS2,MSmassandf}
\begin{equation}
  m(r) = \frac{M r^3}{(r+\ell)^3}\, ,
    \label{ellRS1}
\end{equation}
which gives a particular case of the Fan $\&$  Wang model, investigated in details in Ref.~\cite{Fan:2016hvf}. This model is characterized by strong, order $1/r^2$ corrections to the Schwarzschild solution at infinity. Additionally, the parameter $\ell$ in this model was recently constrained by S$2$ observational data \cite{Cadoni:2022vsn}. 

In the next subsection, we will show that the result \eqref{rhodS2}, which is dictated by an Occam razor argument, can be derived using, again, conformal symmetry arguments.

\subsection{Scalar field description}

The regularization proposed above is obviously not unique. A different choice corresponds to different flows between the $r=0$ and $r = \infty$ points, and to different patterns of the symmetry breaking. The most general solution compatible with the boundary conditions \eqref{rhodS} and \eqref{rhodS1}, and with an analytic behavior at $r \to \infty$ is $\rho\propto (P_n)^\gamma/(P_m)^\delta$, with  $P_n$, $P_m$ polynomials of degrees $n$ and $m$, respectively, and $n\gamma-m\delta= -4$ (to guarantee the scaling \eqref{rhodS1} at great distances). This is a remarkable particular case of the more general formula sourcing regular fractional models
\begin{equation}
    \rho(r) = \frac{\alpha}{4\pi}\frac{\ell^{m-1}}{\left(r^m + \ell^m \right)^{\frac{3}{m}+1}}\, .
    \label{rhogeneralregular}
\end{equation}
Indeed, once used in \cref{generalmetric,MSmassandf}, it gives for $m=1,\, 2,\, 3$, the Fan $\&$ Wang, Bardeen and Hayward black holes, respectively.

To gain further insights into the details of the two CFTs living at $r=0$ and $r=\infty$, let us assume, for simplicity, that there is a regime in which the flow between these two conformal points can be described by a static scalar field $\Phi$, which is sourced by the density $\rho$. In a static and spherically symmetric background, it will satisfy the Poisson equation
\begin{equation}\label{pe}
    \nabla^2 \Phi=4\pi  \rho \, .
\end{equation}
One can now easily  find, using \cref{rhodS,rhodS1}, the asymptotic solutions of \cref{pe} near $r=0$ and $r\to\infty$,
\begin{equation}
    \Phi(r)  \sim  \, \begin{cases} &  r^2/\ell^2 \, , \, \quad \, \,\text{for}\, \quad  r \sim 0\\ &  r^{-2}  \, , \qquad \, \text{for} \, \quad r \to \infty \end{cases} \, ,
    \label{Phibehaviors}
\end{equation}
where we have neglected the constant and $1/r$ terms in the $r\to \infty$ behavior, which are related to the presence of the  mass $M$.
As expected in the flow from $r = \infty$ to $r=0$, the scaling dimension of $\Phi$  changes from its engineering one $\Delta = -2$ to $\Delta = 2$, which is associated with a constant $\rho$.

An important feature of the two conformal points, which is immediately evident in   \cref{Phibehaviors}, is that they are mapped one into the other by the inversion
\begin{equation}
    r \to \frac{\ell^2}{r}\, .
    \label{inversion}
\end{equation}
Discrete symmetries, changing small with large radii, are common in string theory, where they are called $T$-dualities. They have been already used in the past to investigate nonsingular black holes \cite{Smailagic:2003hm, Nicolini:2019irw}. 

The inversion can be used  in combination with translations to produce special conformal transformations, which, together with dilatations and translations, generate the  conformal group (isomorphic to the $\text{SL}(2, \mathds{R})$ group) realized here in one dimension  as 
\begin{equation}
    r \to \omega \, r,\quad r \to \frac{r}{1-\nu r}\,\quad r \to r+\sigma
    \label{ConfTransf}
\end{equation}
where $\omega$, $\nu$, $\sigma$ are the group parameters \footnote{We stress again that a different parametrization of the radial coordinate would simply imply a different realization of the transformations \eqref{ConfTransf}, with $r$ replaced  by a particular function determined by the coordinate transformation (see the discussion in \cref{sec:nonsingular}).}.

A generic flow, for instance the one described by \cref{rhogeneralregular},  will preserve neither the  scaling behavior for $\Phi$  nor the symmetry under inversion. However, we can select a particularly symmetric profile for $\rho$, such that the solution for $\Phi$ preserves at least  part of the conformal symmetries, in particular the scaling with $\Delta = 2$ attained in the $r=0$   conformal point.
One can show that this  happens if we choose the simple profile for $\rho$  given by \cref{rhodS2}. Integrating the Poisson equation \eqref{pe}, we get
\begin{equation}
    \Phi(r) = \frac{\alpha \,}{6 \, \ell^2}\frac{r^2}{(r+\ell)^2}\, .
    \label{potentialregular}
\end{equation}
One can now check that the field $\Phi$ \eqref{potentialregular} transforms as 
\begin{equation}
    \Phi \to \omega^2 \, \Phi 
    \label{scalingtransl}
\end{equation}
under a conformal  transformation of the form
\begin{equation}
    r \to  \omega \frac{r}{1-\nu r}\, ,
    \label{GLtransform}
\end{equation}
with  $\omega \equiv 1+\nu \ell$, which represents the composition of a dilatation  and a special conformal transformation.\\

It is important to notice that \cref{potentialregular} does not arise as the Newtonian limit of the full GR solution with the mass function \eqref{ellRS1}. The EOS \eqref{EoS}, indeed, prevents the weak-field limit from being performed together with the usual nonrelativistic limit, and a Newtonian fluid, with $\rho \gg p_\parallel$, $p_\perp$, from being considered. We can still perform a weak field limit, which gives the Poisson equation, sourced however by the \emph{active mass} $\rho + p_\parallel + 2 p_\perp$. Using \cref{EoS,pperp}, together with the profile \eqref{rhogeneralregular} (with $m=1$) yields the potential $\tilde \Phi = -G M r^2/(r+\ell)^3$.    

\section{Nonlinear Electrodynamics}
\label{sec:nle}

It is interesting to note that the large scale behavior $r^{-4}$ of \cref{rhogeneralregular}, and the related scale invariance, could be explained in terms of the embedding of these regular models as solutions of GR coupled with nonlinear electrodynamics \cite{Bronnikov:2000vy}. The action for such theory is
\begin{equation}
    \mathcal{S} = \int \dd^4 x \, \sqrt{-g}\, \left[\frac{R}{16\pi G}-\mathscr{L}(\mathcal{F}) \right]\, ,
\end{equation}
where $R$ is the Ricci scalar, while $\mathcal{F} = \frac{1}{4}F^{\mu\nu} F_{\mu\nu}$ is the trace of the electromagnetic potential. $\mathscr{L}$ is, in general, a nonlinear function of $\mathcal{F}$. Maxwell's theory is of course recovered in the linear case $\mathscr{L} \propto \mathcal{F}$. If we compute the stress-energy tensor related to $\mathscr{L}$, we see that it naturally satisfies the EOS \eqref{EoS} and that $\rho(r) = \mathscr{L}(\mathcal{F})$. We can now combine this with \cref{rhogeneralregular} and the magnetic monopole solution of Maxwell's equations
\begin{equation}
    \mathcal{F} = \frac{q_\text{m}^2}{2 r^4}\, ,
    \label{FMaxwell}
\end{equation}
which gives the lagrangian
\begin{equation}
    \mathscr{L}\left(\mathcal{F}\right) = \frac{\alpha}{4\pi}\frac{\mathcal{F}}{\left[\ell^\beta \mathcal{F}^{\beta/4} + 2^{-\beta/4}q_\text{m}^{\beta/2}  \right]^{4/\beta}}\, .
    \label{LFregularmodels}
\end{equation}
We see now that the particular large-scale conformal scaling $r^{-4}$ can be explained by the fact that the subclass of models described by \cref{LFregularmodels} reduces to the standard Maxwell theory in the weak field limit $\mathcal{F}\to 0$, which is also conformally invariant. 

\section{Summary and Outlook}
\label{sec:concl}

One of the most unsatisfactory aspect of nonsingular black-hole solutions is that, although we can generate them using anisotropic fluids as sources, their physical origin in terms of elementary fields  is mostly unknown. This is particularly true if one considers nonsingular black holes in which the deformations from the usual Schwarzschild solution have superplanckian origin \cite{Cadoni:2022chn,Cadoni:2022vsn}. An unpleasant consequence of this lack of knowledge is the existence of a large number of solutions. Although it is possible that, in the near future, astrophysical and  gravitational waves data may be used to select/exclude models \cite{Cadoni:2022chn,Cadoni:2022vsn}, some theoretical guiding principle is more than welcome.

It is likely that these difficulties are indicating that the microscopic origin of this kind of solutions cannot be found in a GR framework or its extensions (see Ref.~\cite{Knorr:2022kqp}), which allows only for a coarse-grained description in terms of anisotropic fluids. For this reason, it is important to look at general guiding principles, like symmetries, which are expected to underpin the classical GR description.

In this paper, we have adopted this philosophy to constrain the broad class of nonsingular black-hole models with a dS core. 
We have used conformal symmetries,  which  are believed to be a crucial ingredient of any quantum theory of gravity,  as a selecting principle to single out the physically relevant nonsingular black-hole solution. 

We have found that the conformal symmetry selects a particular case of the Fan $\&$ Wang metric, which has been recently investigated and constrained using data of the orbits of the S$2$ star around  the SgrA$^\ast$ black hole.  

Obviously, the use of conformal symmetry to select solutions is only a first step. Understanding the microphysics from which these symmetries originate is the next important task. 

\bibliography{refs}
\end{document}